\documentclass[aps,superscriptaddress,showpacs,amsmath,amssymb,amsfonts,altaffillsymbol,twocolumn,notitlepage,prx]{revtex4-1}

\usepackage{graphicx}
\usepackage{float}
\usepackage{dcolumn}
\usepackage{bm}
\usepackage{color}
\usepackage{hyperref}

\graphicspath{{./figures/}}

\begin{document}

\title{Memory behavior of a randomly driven model glass}

\author{Roni Chatterjee}%
\affiliation{Tata Institute of Fundamental Research, 36/P, Gopanpally Village, Serilingampally Mandal,
Ranga Reddy District, Hyderabad 500046, Telangana, India}%

\author{Smarajit Karmakar}%
\affiliation{Tata Institute of Fundamental Research, 36/P, Gopanpally Village, Serilingampally Mandal,
Ranga Reddy District, Hyderabad 500046, Telangana, India}%

\author{Muhittin Mungan}
\affiliation{Institute for Biological Physics, University of Cologne, Z{\"u}lpicher Stra{\ss}e 77, K{\"o}ln, Germany}

\author{Damien Vandembroucq}%
\affiliation{PMMH, CNRS, ESPCI Paris, Universit\'e PSL, Sorbonne Universit\'e, Universit\'e Paris Cit\'e, France}%

\date{\today}

\begin{abstract}
We investigate by atomistic simulations the memory behavior a model glass subjected  to random driving protocols. The training consists of a random walk of forward and/or backward shearing sequences bounded by a maximal shear strain of absolute value {$\gamma_T$}. We show that such a stochastic training protocol is able to record the training amplitude. Different read-out protocols are also tested and are shown to be able to retrieve the training amplitude. We then emphasize the tensorial character of the memory encoded in the glass sample and then characterize the anisotropic mechanical behavior of the trained samples.
\end{abstract}

\maketitle

\section{Introduction}

Disordered materials retain traces of their past, and their behavior—whether they are glass, gel, granular media, fabrics, or knits—depends on their thermal and mechanical history. It is natural to expect that the associated information is hidden within the disordered structure. But is it possible to access this information? Can we record and retrieve the memory of disordered matter? These questions are garnering increasing interest among soft matter physicists, as evidenced by a surge in recent research highlighted in reviews on the subject~\cite{Keim-RMP19,Paulsen-Keim-ARCMP25}.  Notably, various experimental systems have demonstrated that mechanical annealing through cyclic shear allows us to record and later read the amplitude and direction of the applied oscillatory shear \cite{mukherji2019strength,ghosh2022coupled,keim2020global}.

Cyclic protocols are frequently employed to assess the fatigue resistance of materials \cite{parmar2019strain, maity2024fatigue, bhaumik2021role, chatterjee2025effect, fiocco2013oscillatory, leishangthem2017yielding, yeh2020glass, chatterjee2024role}. However, in real-world applications, most materials encounter fluctuating mechanical loads due to environmental variations. Granular soils, for example, are affected by contraction and dilation as well as changes in friction coefficients caused by daily or seasonal fluctuations in temperature and humidity \cite{Jagla-Ferrero-Soil25}. Similarly, clothing experiences variable mechanical loads that depend on the wearer's morphology and physical activity. In a biological context, the extracellular matrix is subjected to forces exerted by migrating cells \cite{Adar-Joanny-PRL25}.

{In this study,} we are interested in the mechanical imprint that a {fluctuating} mechanical environment can induce on a disordered material. This study {builds on previous work on subjecting a mesoscopic elastoplastic model} to random loading~\cite{Mungan-PRL25}. It also echoes recent work on the mechanical memory of active glasses~\cite{Karmakar-NatPhys25,Sastry-NatPhys25, agoritsas2024memory}, {in which the driving protocols have inherent stochasticity}.

More specifically, we use atomistic simulations to study a model glass under random loading; we show that it is possible to readout the maximum deformation amplitude using a dedicated reading protocol. {These findings are consistent with our earlier results obtained from simulations of a mesoscale model of a glass and thus validate these.}   We then focus on the mechanical behavior of glasses  {annealed by random shearing. We find that their stress-strain response is anisotropic and polarized. We show that the memory of random training also encodes the plane of applied shear and thus cannot be read-out in other shear directions}~\cite{Adhikari-Sharma-Karmakar-PRL25}. Performing monotonic shear tests on the trained glass, we recover a Bauschinger-like \cite{patinet2020origin,karmakar2010plasticity} behavior in the training direction.   

This article is organized as follows. In Section \ref{sec:training_protocol}, we provide a brief overview of our model glass, detailing its preparation and the mechanical training protocols used. Next, in Section \ref{sec:read-out}, we discuss the results of applying various read-out protocols to the trained glass in order to determine the conditions under which the memory of the training amplitude can be recovered. Section \ref{sec:anisotropy} focuses on the mechanical response of the trained glass, particularly examining the stress-strain response to different modes of shear strain applied. We conclude with a discussion of our findings in Section \ref{sec:discussion}.     

\section{Preparation and Mechanical Training of Model Glasses}
\label{sec:training_protocol}

{\em Sample Preparation --} We prepared two-dimensional (2D) Kob-Andersen model glasses with sizes $N=400,\; 1000,\; 2000$. These glass samples were equilibrated at a parent temperature of $T_p = 1.0$, and their inherent structures were obtained through an instantaneous quench at $T = 0$. For each size, we prepared $32$ independent samples. We then subjected these inherent structures to a protocol of athermal quasistatic shear (AQS) \cite{maloney2006amorphous}. Additional details on the numerical implementation can be found in Appendix~\ref{sec:appendix-numerical-details}.



\begin{figure*}
\centering
  \includegraphics[width=0.95\textwidth,height=0.5\textwidth]{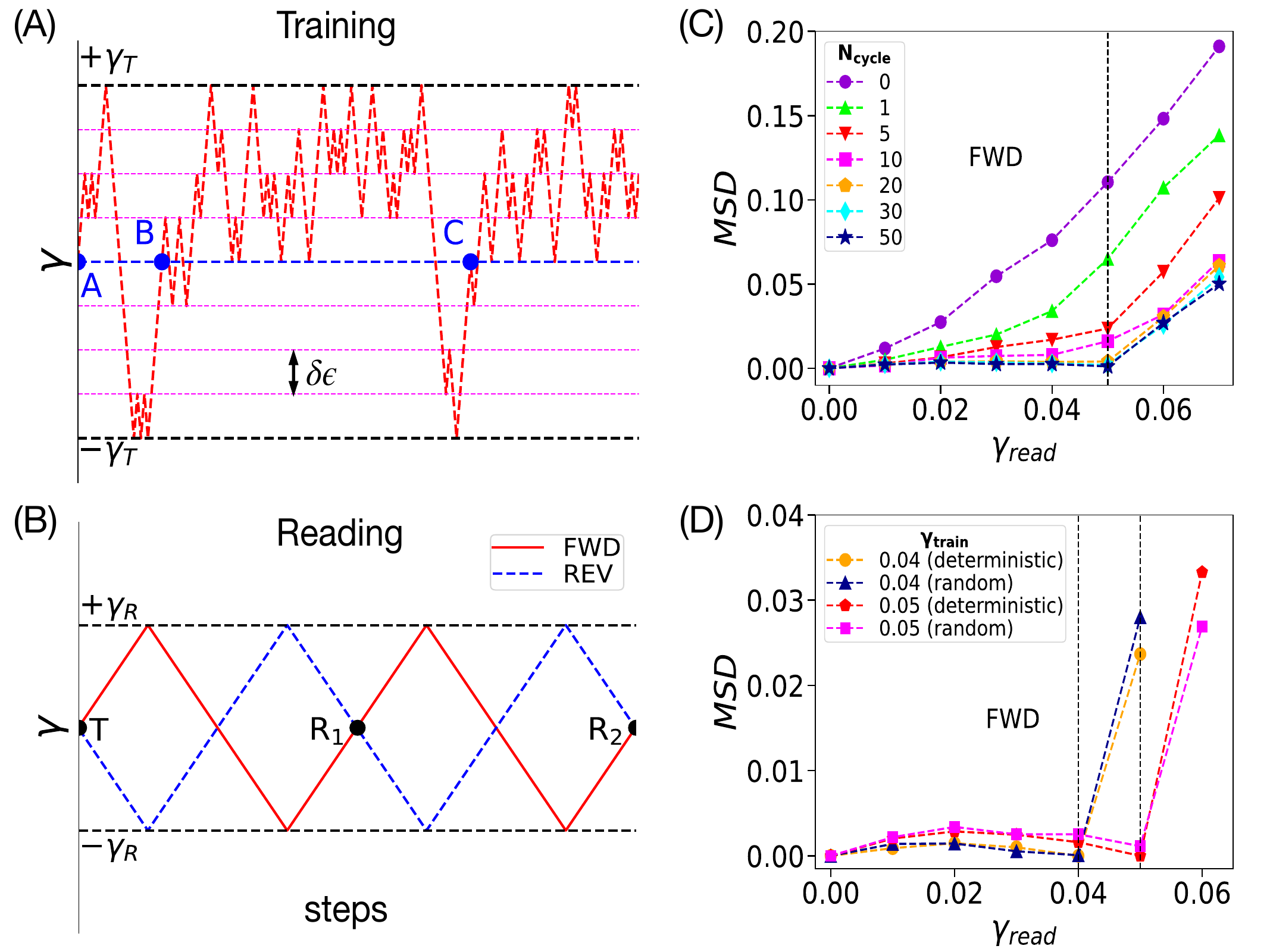}
  \caption{{\bf (A)} Illustration of the random driving protocol as a random walk along the strain axis with fixed strain step size $\delta \epsilon$ (indicated by the dashed red lines) and reflecting boundaries at $\pm \gamma_T$ (horizontal black dashed lines). A random driving cycle consists out of a first passage from the initial state $A$ to one of the reflecting boundaries, which is followed by a first passage to the opposite boundary and a subsequent (first) return to zero strain. The training by random cycles involves applications of multiple cycles with the same sequence of boundary visits as the first cycle, leading to the ``trained" glass $T$. Without lack of generality these cycles can be rectified such that the first boundary visited always establishes the direction of positive shear. {\bf (B)} Read-outs from $T$ are performed by applying one or two cycles of (deterministic) cyclic shear with read-out amplitude $\gamma_R$ that either have the same sense of driving as the random driving (FWD), or are out of phase with respect to it (REV). The states reached at the end of the first and second read-out cycles are denoted by $R_1$ and $R_2$, respectively. {\bf (C)} The result of a read-out obtained by applying to $T$ FWD read-out cycles with different strain read-out amplitudes $\gamma_R$ and measuring the mean square displacement (MSD) between the particle configurations of $T$ and $R_1$. Shown are the result for a system with $N = 2000$ particles, trained at $\gamma_T = 0.05$ for various numbers of training cycles as indicated in the figure. The FWD MSD read-out curve develops a kink at a read-out amplitude matching the training amplitude (vertical dashed lines). Already after $30$ training cycles, this kink has developed into a local minimum so that the MSD vanishes when training and read-out strains match.  {\bf (D)} Comparison of FWD memory signals from samples trained either under a deterministic and or random cyclic shear protocol for $N = 2000$ and training amplitudes $\gamma_T = 0.04$ and $0.05$. For each training amplitude shown, annealing by deterministic and random cycling leads to nearly indistinguishable MSD read-outs.    
  }
  \label{intro-fig}
\end{figure*}

{\em Random Driving Protocol --} As in \cite{Mungan-PRL25}, the driving by random shearing is implemented as a random walk (RW) along the strain axis with reflecting boundaries at $\pm \gamma_T$ and using a fixed RW strain step size of $\delta\epsilon$, as illustrated in Fig.\ref{intro-fig}(A). A single training cycle is defined by a random walk and its first passages to the reflecting boundaries at $\pm \gamma_T$ as follows: starting at zero strain, the RW hits one of the boundries for the first time, then reaches the opposite boundary and finally returns from there to zero strain for the first time, as illustrated in Fig.\ref{intro-fig}(A), with a random cycle starting at $A$ and ending in $B$. Since the boundary hit first was at $+\gamma_T$, the second driving cycle will be formed by the first passage from $B$ to $+\gamma_T$, then to $-\gamma_T$ and finally the first subsequent return to zero strain at C. Since the applied strain path is random, initially either the $+\gamma_T$ or $-\gamma_T$ boundary will be hit first, establishing a sense of the cycle, which every subsequent random cycle will then follow, as explained above. Observe that by isotropy of the positive and negative strain directions, without loss of generality, we can ``rectify" the driving cycles for which the $-\gamma_T$ boundary is hit first by inverting the positive direction of shear strain (and stress). Thus, after ``rectification", all random driving cycles hit the $+\gamma_T$ boundary first.  

We trained $32$ samples of system size $N = 2000$ at strain amplitudes $\gamma_{T}  = 0.04, 0.05$. For this system size, the yield strain had been previously determined to be $\gamma_{yield} \approx 0.065$ \cite{chatterjee2025effect}. We considered training by deterministic cyclic shear as well as for the random cycling described above. For the samples trained by random cycling, the random walk step size used was  $\delta\epsilon = 0.01$.
{Our training protocol consisted of applying up to $N_{\rm cycle} = 100$ cycles.} We found that in the case of random driving, memory signals were already well-established after $20$ training cycles.




{\em Read-Out Protocol --} After subjecting the prepared glasses to a number of random driving cycles, denoted as $N_{\text{cycle}}$, at a specific ``training" strain $\gamma_T$, we obtain a trained glass sample referred to as $T$. The read-out process involves applying deterministic cyclic shear at a variable read-out amplitude $ \gamma_R $, as illustrated in Fig. \ref{intro-fig}(B). This cyclic read-out can be either ``in phase" or ``out of phase" with respect to the direction of the (rectified) training that occurred through random driving. In the ``in-phase" condition, the applied cyclic shear follows the sequence: $ 0 \to +\gamma_R \to -\gamma_R \to 0 $. Conversely, in the ``out-of-phase" condition, the applied cyclic shear follows the sequence: $ 0 \to -\gamma_R \to +\gamma_R \to 0 $. We refer to these two directions of read-out as ``forward (FWD)" and ``reverse (REV)", respectively. Specifically, the read-out process consists of applying two cycles with the same sense and read-out amplitude. We label the states of the glass at the end of the first and second cycles as $R_1$ and $R_2$, respectively. 


In order to compare the glasses reached after training with those after the various read-outs,  we compute the mean squared displacement (MSD) between the particle configuration of the trained glass $T$ and the state $R_1$. We shall refer to these as the FWD or REV read-outs, depending on the sense of the read-out cycle, as explained above and illustrated in Fig.~\ref{intro-fig}(B). In addition, we also record  the MSD between the particle configurations $R_1$ and $R_2$ that are reached at the end of the first and second read-out cycles, and refer to these as the FWD-FWD and REV-REV read-outs. All MSD read-out results shown in the following are averages over the $32$ trained samples.

{

\section{Mechanical memory read-out in different shear directions}
\label{sec:read-out}

{Until} now, the classical method for recovering the training amplitude of a sample that has been trained involves applying a series of cycles with increasing read-out amplitudes. During this process, the similarity between the configurations obtained after training and those produced during the read-out cycle is assessed using the Mean Square Displacement (MSD)~\cite{Keim-RMP19}. Experiments have shown that when training is done through deterministic cyclic shearing at an amplitude $\gamma_T$ that is below the yielding point, the maximum similarity is achieved when the read-out amplitude $\gamma_R$ matches the training amplitude $\gamma_T$ \cite{keim2020global, Benson-PRE21, Cloitre-memory-JRheo22, Shohat-PNAS22, Keim-Medina-SciAdv22}. This type of read-out is referred to as ``sequential."  Alternatively, in numerical simulations, instead of applying a sequence of read-out cycles with increasing amplitudes, it is possible to perform the read-out cycles at different amplitudes in parallel across copies of the trained state $T$. This method is known as ``parallel" read-out. The formation of memory under deterministic cyclic shearing combined with parallel read-outs has also been confirmed numerically \cite{Keim-PRL11, Fiocco-PRL14, Adhikari-Sastry-EPJE18, Regev-PRE21, Kumar-PRE25}.

Note that such read-out protocols implicitly assume the knowledge of the sense of driving, i.e. direction of the shear cycle. In the case of random driving, we used this knowledge to ``rectify" the sense of driving cycles.  In Ref.~\cite{Kumar-PRE25} some of us developed read-out protocols designed to extract the shear direction of the training protocol in addition to its amplitude. Moreover, in Ref.~\cite{Mungan-PRL25} we showed that the  FWD-FWD and REV-REV read-out protocols are direction-insensitive and can reveal the training amplitude without having to know the sense of driving in advance. The read-out protocols considered in these works simulated a mesoscale model of an amorphous solid. Here we are interested to apply the single FWD/REV and two-cycle FWD-FWD/REV-REV read-outs of Ref.~\cite{Mungan-PRL25} to an atomistic model of a glass former.




We start by investigating the formation of mechanical memory by systematically examining how the memory signal develops with increasing training cycles ($N_{\text{cycle}}$). Fig.\ref{intro-fig}(C) shows the evolution of this signal for the FWD read-out protocol. These measurements were conducted on the systems of size $N = 2000$ trained at a strain amplitude of $\gamma_T = 0.05$. Our findings demonstrate that untrained samples ($N_{\text{cycle}} = 0$) show no memory effect.
As the number of training cycles increases, a memory signal emerges and progressively strengthens, reflecting the gradual encoding of deformation history into the material's microstructure. This is apparent from the emergence of a kink in the MSD read-outs after $N_{\text{cycle}} = 5$ around read-out amplitudes matching the training amplitude, as indicated by the dashed vertical line. The memory signal appears to saturate after approximately $N_{\text{cycle}} = 20$, suggesting that the system has reached a stable trained state. Beyond this point, additional cycles do not enhance the encoding of amplitude memory. For the system size and random walk step size considered, $20$ training cycles are sufficient to establish a well-defined and saturated mechanical memory in the sample.

\begin{figure}[t!]
  \includegraphics[width=0.47\textwidth,height=0.35\textwidth]{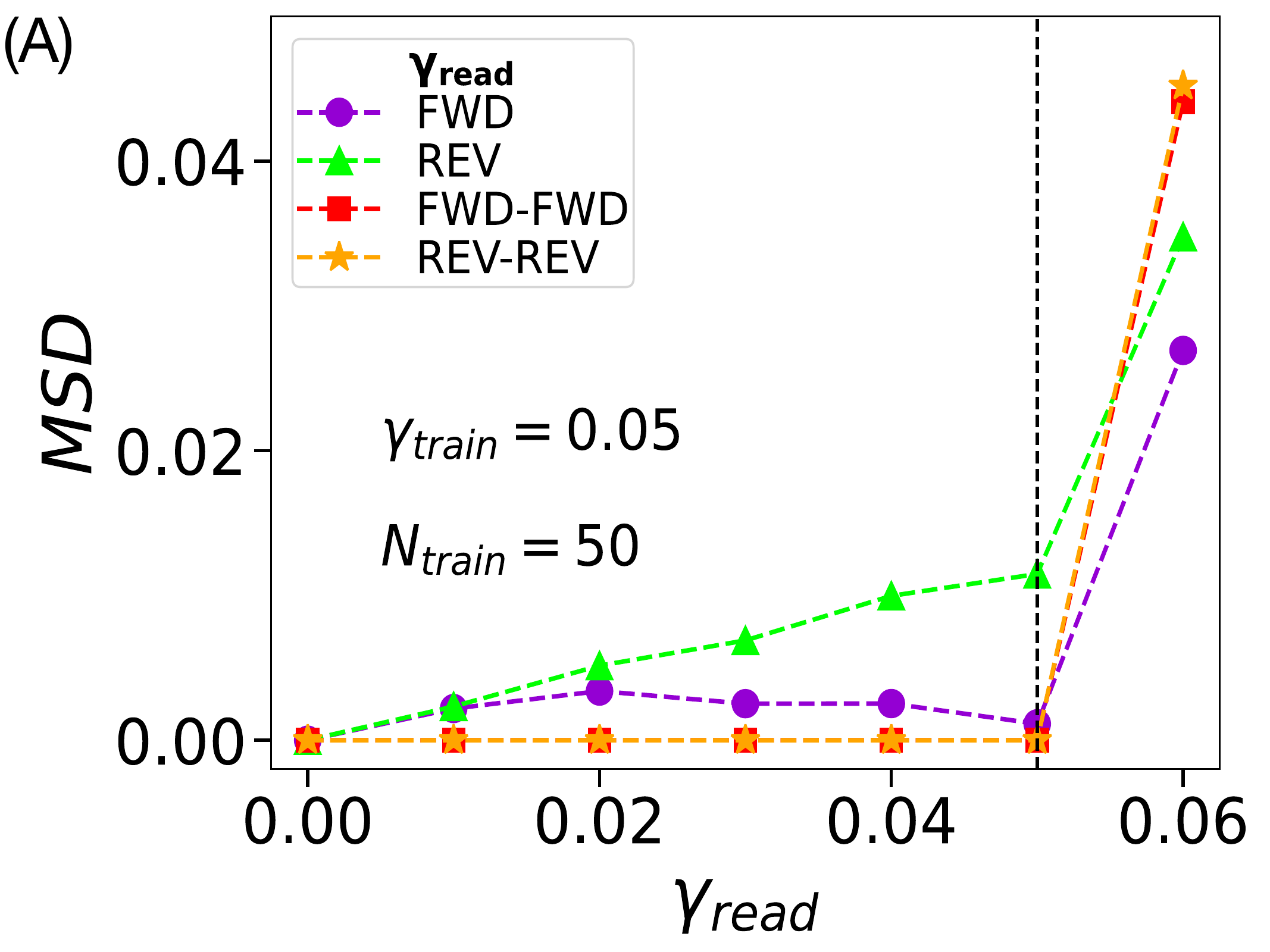}
  \includegraphics[width=0.47\textwidth,height=0.35\textwidth]{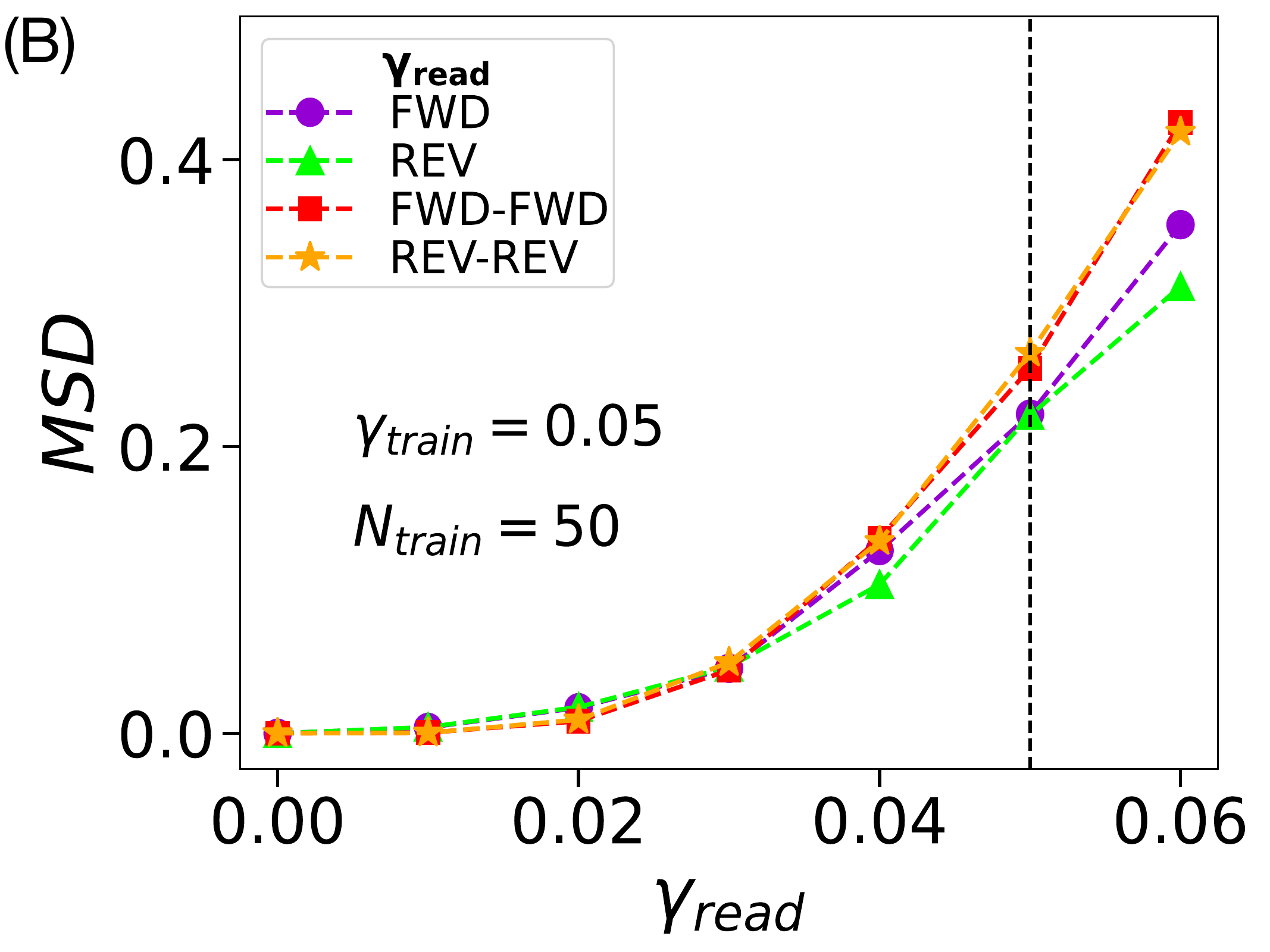}
  \caption{{\bf (A)} The results of the four read-out protocols FWD, REV, FWD-FWD and REV-REV, applied to a glass subject to random cycling over $50$ training cycles at amplitude $\gamma_T = 0.05$. The FWD read-out shows a local minimum where the read-out and training amplitude match, as indicated by the black dashed vertical line. In the REV read-outs, the MSD increases with increasing $\gamma_R$ exhibiting a change of growth rate around $\gamma_R \approx \gamma_T$. Note that in the absence of the knowledge of the positive sense of shearing cyclic shear applied to a sample will result either in a FWD or REV read-out, which are qualitatively different. Conversely, the FWD-FWD and REV-REV exhibit nearly indistinguishable read-out responses and hence are independent on the orientation of the shear plane. Moreover, for the training amplitude shown the 2-cycle read-out signal remains close to zero rising sharply only when $\gamma_R > \gamma_T$.
  {\bf (B)}. Read-outs of a sample trained under random cycling by \textbf{\textit{simple shear}}, but where read-outs are performed by deterministic cycles of \textbf{\textit{pure shear}}, corresponding to compression and elongation of the sample along the horizontal and vertical axis such that the area remains constant. All four read-out protocols exhibit very similar MSD curves and moreover show no memory of the amplitude of training under simple strain.   
  }
   \label{panel_2}
\end{figure}

We next perform a comparative analysis of the FWD memory signal resulting from deterministic versus random training protocols. Again, $32$ {independently prepared samples} of size $N = 2000$ were subjected to a random training protocol at strain amplitudes of $\gamma_T = 0.04$ and $0.05$,  using a RW step size of $\delta\epsilon = 0.01$. In addition, we took the same untrained samples and subjected them to a deterministic cyclic shear protocol with the same amplitudes of $\gamma_T = 0.04$ and $0.05$. Fig.\ref{intro-fig}(D) presents the result of  FWD read-outs for these differently trained samples. The results indicate that the magnitude and character of the memory signal show negligible dependence on the training method. This suggests that the establishment of mechanical memory is a robust phenomenon, largely independent of whether the training deformation is applied deterministically or stochastically.


 In Fig.~\ref{panel_2}(A) we have plotted the FWD, REV, FWD-FWD and REV-REV read-out responses subsequent to random training at $\gamma_T = 0.05$. As already discussed before, the FWD read-out curve exhibits a  minimum at {$\gamma_{R} = \gamma_{T}$}. This indicates that within one cycle the particles rearrange themselves in a way so that they are coming back very close to their previous locations. In case of the REV response, there is no such indication of memory. Instead,  we observe that the MSD curve rises nearly monotonously, changes its slope significantly when {$\gamma_{R} \approx \gamma_{T}$}. Note again that in the absence of  knowledge of the direction of shear (but knowledge of the shear plane), a cyclic read-out cycle applied to the trained sample would have resulted either in a FWD or REV read-out, hence in a direction-dependent response. However, this directional dependence disappears, and directional independence is gained, when we consider the two-cycle read-out protocols FWD-FWD and REV-REV. As seen in Fig.~\ref{panel_2}(A), the FWD-FWD and REV-REV read-out curves are statistically indistinguishable. Moreover, the read-out MSD is zero to machine precision for read-out amplitudes less than training amplitude i.e. {$\gamma_{R} \leq \gamma_{T}$}. These findings are consistent with our earlier results obtained from simulations of a mesoscale elastoplastic model of an amorphous solid subject to training by random cycling \cite{Mungan-PRL25}.

Note that the training and read-out have been done by the application of simple shear, i.e. deformation of the square-shaped sample into  a parallelogram. An alternative shear deformation is via one of the two pure shear modes. Here we will consider the shear mode where one side of the square-shaped sample is compressed, while the other is elongated so that a rectangle of the same area is obtained. In the following, we will refer to this  mode of pure shear as push-pull.

We next probe the anisotropy of the encoded memory, by asking whether a memory signal persists when a sample trained by simple shear is subject to push-pull shear read-outs along the same shear plane. 
Samples trained under both deterministic and random driving under simple shear at $\gamma_T = 0.05$, were subsequently subjected to a push-pull reading protocol.
As shown in Fig.~\ref{panel_2}(B), the mechanically annealed material exhibits an isotropic mechanical response in this orthogonal reading geometry, without any memory signal. We obtained a qualitatively similar result for training by deterministic cyclic shear (not shown). These findings  demonstrate that the mechanical memory is highly directional and is only accessible when the material is probed along the same shear plane in which the training was performed.

\begin{figure*}
\centering
  \includegraphics[width=1.0\textwidth,height=0.325\textwidth]{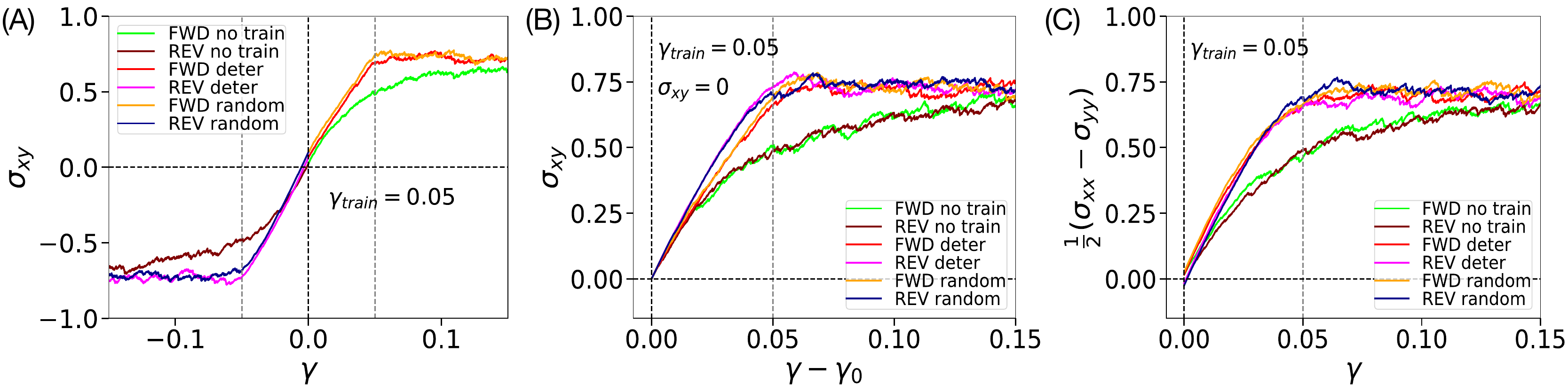}
  \caption{Mechanical stress-strain response after training. {\bf (A)} stress-strain curves
  obtained  in the forward and reverse direction {\it i)} for the pristine glass, {\it ii)} after deterministic cyclic training at $\gamma_T=0.05$ and, {\it iii)} after random training at $\gamma_T=0.05$. After mechanical annealing the trained glasses are stiffer and harder than the pristine glass. {\bf (B)} Mechanical response of the same samples after  prior unloading to a stress free state. After symmetrization of the reverse response, the pristine glass shows perfect symmetry between forward and reverse loading while the trained glasses  show a significant asymmetry: after training the samples have become harder in the reverse direction than in the forward direction. {\bf (C)} Mechanical response in the pure shear $xx-yy$ direction. Again,  
  the pristine glass shows perfect symmetry between forward and reverse loading. This symmetry is also recovered for the trained sample.}
  \label{fig:after-training}
\end{figure*}

\section{Anisotropic mechanical behavior of the trained glass}
\label{sec:anisotropy}

Here we characterize the mechanical behavior of the trained glass. It
is actually of interest to distinguish between two different initial
configurations for the mechanical tests to be performed: i) the
trained state $T$ obtained after random (or deterministic) training ;
ii) the stress relaxed configuration $T_0$ obtained from $T$ after the
necessary unloading.

{We conducted simple shear tests in both forward and reverse shear directions using the specified configurations. The results are presented in Fig.~\ref{fig:after-training}A, which illustrates the trained states $T$ obtained with an amplitude of $\gamma_T = 0.05$. For reference, we also include the stress-strain curves of the pristine glass. The results for both random driving and deterministic cycling driving are comparable. In both scenarios, mechanical annealing increases the stiffness and hardness of the glass. The effective shear modulus, indicated by the slope at the origin, is slightly higher, and the elastic range is significantly expanded compared to that of the pristine glass. In addition to this hardening effect, the trained glass exhibits a slight polarization. The opening of the hysteresis cycle results in a non-zero residual shear stress post-training. Specifically, we measured $\sigma_{xy}^{\text{rand}} = 0.1087 \pm 0.0821$ for random training, and $\sigma_{xy}^{\text{det}} = 0.0635 \pm 0.0901 $ for deterministic cycling training at $\gamma_T = 0.05$. These values should be compared to the typical measurement from poorly annealed pristine glass, $\sigma_{xy}^{0} = 0.02247 \pm 0.0773$, which is influenced by sample-to-sample finite-size fluctuations. Recent studies have also considered fluctuations in local strain or stress of pristine samples as indicators of the degree of annealing \cite{pollard2022yielding}.
.}

{In Fig.~\ref{fig:after-training}B, we show the mechanical response obtained when starting from the stress-free states $T_0$ (obtained after unloading from the trained states $T$). The responses in the reverse loading directions have been reflected around the origin for easier comparison with their counterparts in the forward loading direction.  As a reference, we use the response of the pristine glass. As anticipated, the pristine case demonstrates complete symmetry between the forward and reverse loading. The stress-strain curves in this scenario are statistically indistinguishable.  
}

{In contrast, we observe for both trained glasses (randomly or cyclically) a significant asymmetry between the two loading senses. The forward response is systematically softer than the reverse response and we recover here a Bauschinger-like effect~\cite{patinet2020origin,karmakar2010plasticity}. Note that the present results are perfectly consistent with the recent findings of Kumar {\it et al}~\cite{Kumar-PRE25} obtained in the framework of a quenched mesoscopic elastoplastic model, showing that in the trained state, the forward plastic strengths are systematically lower than the reverse ones. Mechanical annealing thus induces a mechanical polarization effect whose sign is controlled by that of the last limiting strain $\pm \gamma_T$ visited in the course of the mechanical history.   
}

{In addition to the simple shear responses, we present the stress-strain curves along the pure shear direction in Fig.~\ref{fig:after-training}C. The results are reflected onto the positive quadrant to allow for a comparison between forward and reverse loading. Similar to the findings in the simple shear direction, we observe a significant hardening effect compared to the pristine glass. However, unlike the polarization effect seen in the simple shear direction, the forward and reverse loading curves overlap in their fluctuating behavior.
}

{The mechanical training in the $xy$ direction thus induced an isotropic effect of hardening in the material but left in addition a significant polarization along the $xy$ direction.}

\section{Discussion}
\label{sec:discussion}

In this work, we have investigated the memory behavior of a model glass subjected to random driving using atomistic simulations. By introducing a stochastic training protocol based on a random walk in strain space, we demonstrated that the material retains a clear mechanical memory of the training amplitude. Different read-out protocols were presented. In particular, the 2-cycle read-out protocol defined in Ref.~\cite{MKPV-PRL25} gives a robust access to the maximum amplitude of the training since it is fully independent on the sense of the last shear steps.  

Our results reveal that memory formation occurs only below the yielding point; once the system is driven beyond yield, irreversible rearrangements erase any trace of deformation history (see Appendix \ref{sec:above-yielding}). The memory signal emerges gradually with the number of training cycles and eventually saturates, marking the formation of a stable trained state. Importantly, we find that the establishment of memory is robust, as it does not depend on whether the training is applied deterministically or stochastically. Moreover, the encoded memory is highly directional: it is retrievable only along the shear plane in which the material was trained, underscoring its anisotropic character.

Beyond identifying the existence of memory in randomly driven glasses, we characterized the anisotropic mechanical response of trained samples, which bears similarities to the Bauschinger-like effects in plastically deformed solids. These findings extend our understanding of how disordered materials encode their deformation history and provide a framework for probing memory in systems subjected to fluctuating, non-cyclic environments.

Overall, our study highlights the universality and robustness of mechanical memory in glasses and suggests that random driving -— more representative of realistic operating conditions than idealized cyclic loading -— can still leave persistent and retrievable imprints on disordered matter.

\section*{Acknowledgments}
 
SK and RC acknowledge funding by intramural funds at TIFR Hyderabad from the Department of Atomic Energy (DAE) under Project Identification No. RTI 4007. SK acknowledges the Swarna Jayanti Fellowship grants DST/SJF/PSA01/2018-19 and SB/SFJ/2019-20/05 from the Science and Engineering Research Board (SERB) and Department of Science and Technology (DST). SK would like to acknowledge the research support from the MATRICES Grant MTR/2023/000079, funded by SERB. Most of the computations are done using the HPC clusters procured using Swarna Jayanti Fellowship grants DST/SJF/PSA01/2018-19, SB/SFJ/2019-20/05 and Core Research Grant CRG/2019/005373.

\appendix
\section{Numerical Methods}
\label{sec:appendix-numerical-details}

A two dimensional (2D) binary Kob-Andersen model glass \cite{kob1994scaling,kob1995testing,kob1995testing,bruning2008glass} is considered. The mixing ratio of A and B-type particles is $65:35$. The pairwise interaction potential is given by
\begin{equation*}
V_{\alpha\beta}(r)=4\epsilon_{\alpha\beta}\left[\left(\frac{\sigma_{\alpha\beta}}{r}\right)^{12}-\left(\frac{\sigma_{\alpha\beta}}{r}\right)^6+C_0+C_2\left(\frac{r}{\sigma_{\alpha\beta}}\right)^2\right]
\end{equation*}
where $\alpha,\beta \in {A,B}$. The interaction parameters are: $\epsilon_{AB}/\epsilon_{AA} = 1.5, \epsilon_{BB}/\epsilon_{AA} = 0.5$; $\sigma_{AB}/\sigma_{AA} = 0.8, \sigma_{BB}/\sigma_{AA} = 0.88$. The potential is truncated at $r_c = 2.5\sigma_{\alpha\beta}$. The system is simulated at a number density $\rho = 1.2$ with system sizes $N = 400$, $1000$, and $2000$ particles. Molecular dynamics simulations are performed in the NVT ensemble (constant particle number $N$, volume $V$, and temperature $T$) using the Nose–Hoover thermostat in a square box with periodic boundary conditions. The integration time step is $dt = 0.005$. The liquid configurations are equilibrated at parent temperature $T_p = 1.0$. For each size, $32$ independent samples are prepared. Glass states are obtained by an instantaneous quench to $T = 0$ from the equilibrated configurations using conjugate gradient (CG) method. The resulting glass is poorly annealed. These minimised configurations are called Inherent Structures (IS).}

\renewcommand\thefigure{A\arabic{figure}}
\setcounter{figure}{0}

\begin{figure}[t!]
  \includegraphics[width=0.45\textwidth,height=0.35\textwidth]{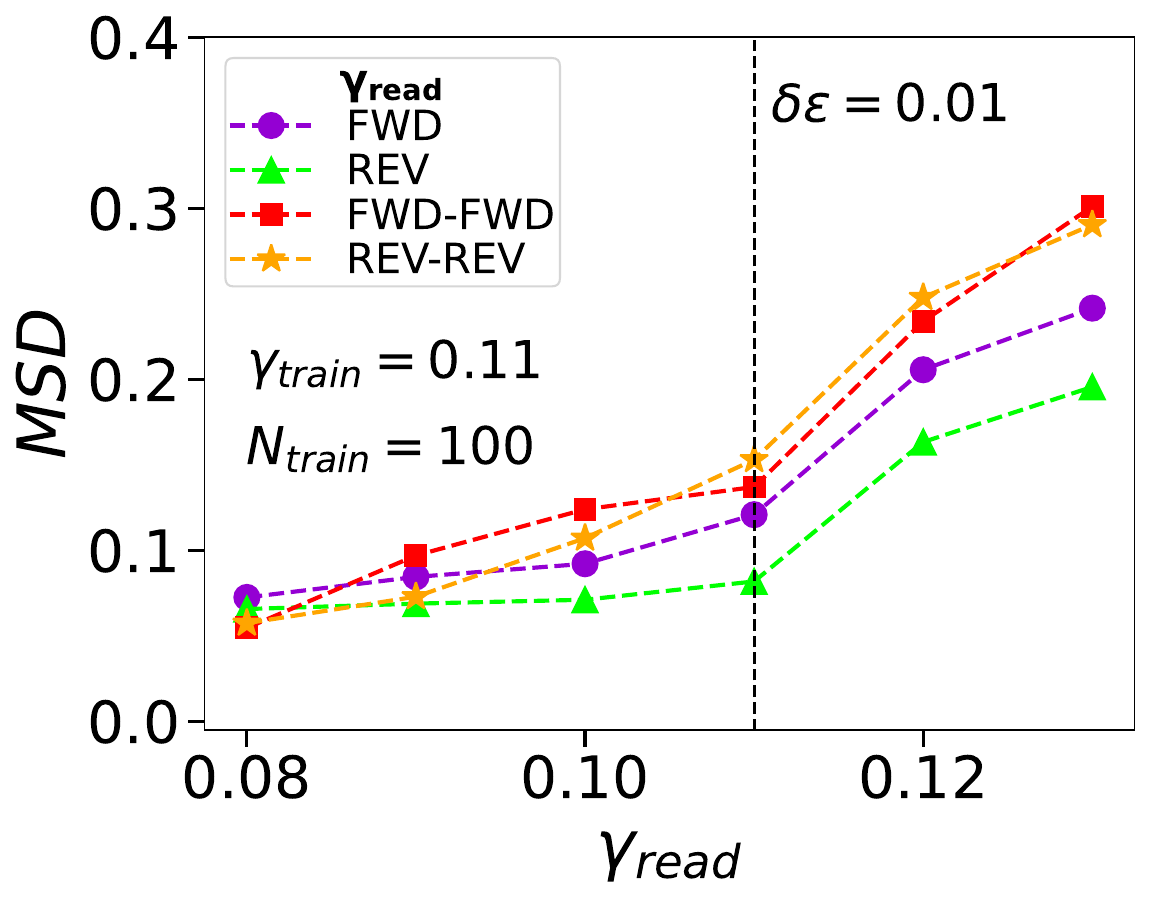}
  \includegraphics[width=0.45\textwidth,height=0.35\textwidth]{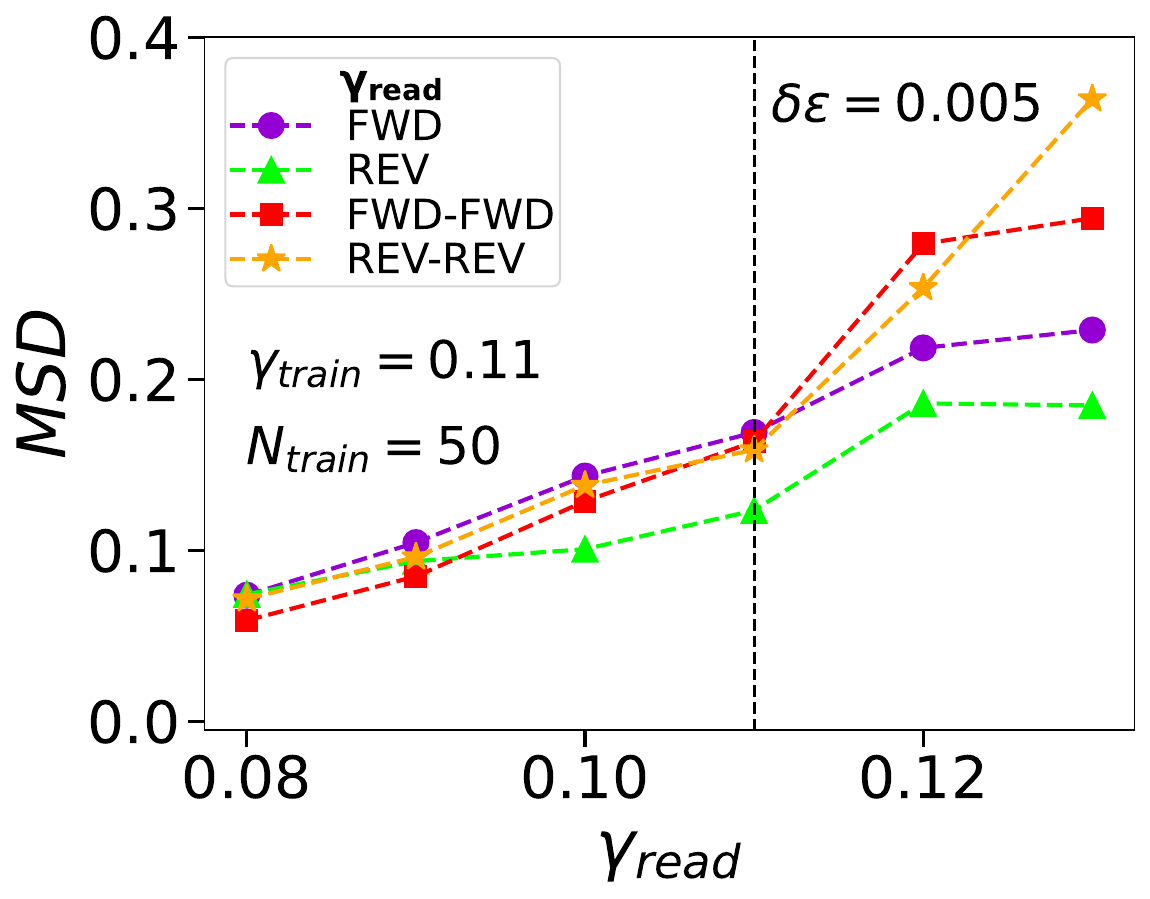}
  \caption{Response of read-out protocols applied to a glass sample trained at $\gamma_T = 0.11$ ($> \gamma_{\text{yield}}\approx 0.09$) for two different random walk step sizes --- $\delta\epsilon = 0.01$ and $0.005$. We don't observe any memory formation at training amplitude beyond yielding.}
   \label{appendix_1}
\end{figure}

The inherent structures were then subjected to a protocol of athermal quasistatic shear (AQS) \cite{maloney2006amorphous}, where the shear strain is changed by a elementary strain unit of  $d\gamma = 2 \times 10^{-4}$ and  Lees–Edwards boundary conditions \cite{lees1972computer} are employed during shearing. Specifically, each elementary straining step consists of two steps: (i) an affine displacements of particles, $r^\prime_x \rightarrow r_x + r_y \times d\gamma, r^\prime_y \rightarrow r_y$; followed by (ii) energy minimisation via a conjugate gradient (CG) method.



\section{Training at amplitudes above yielding}
\label{sec:above-yielding}

We investigated memory formation in samples trained beyond their yielding point. Systems of size $N = 400$ were subjected to a random training protocol at a strain amplitude of $\gamma_{\text{train}} = 0.11$ ($> \gamma_{\text{yield}}\approx 0.09$). This protocol was applied using two different step size --- $\delta\epsilon = 0.01$ and $\delta\epsilon = 0.005$, as shown in Fig.~\ref{appendix_1}.

Our results demonstrate a fundamental difference in memory behavior in post-yielding case. In contrast to the robust memory formation observed below the yield strain, training at $\gamma_{\text{train}} > \gamma_{\text{yield}}$ results in a complete absence of memory. This indicates that the irreversible structural rearrangements associated with yielding erase any specific deformation history imparted during training, preventing the establishment of a stable mechanical memory.

\bibliographystyle{apsrev4-1}
\bibliography{memory}

\end{document}